\documentclass[runningheads]{llncs}
\usepackage[T1]{fontenc}
\usepackage{graphicx,verbatim}
\usepackage{multirow}
\usepackage{tabularx}
\usepackage{amssymb}

\usepackage{marvosym}
\usepackage{amsmath}

\begin{document}

\title{Neural Proteomics Fields for Super-resolved Spatial Proteomics Prediction}

\author{
Bokai Zhao$^\text{1,2,3,4}$, 
Weiyang Shi$^\text{2,3}$, 
Hanqing Chao$^\text{4,5}$, 
Zijiang Yang$^\text{4}$, 
Yiyang Zhang$^\text{2,3}$, 
Ming Song$^\text{1,2,3}$, 
and Tianzi Jiang$^\text{1,2,3}$\textsuperscript{\Letter} 
}
\authorrunning{Zhao et al.}
\institute{$^\text{1}$School of Artificial Intelligence, University of Chinese Academy of Sciences. \\
$^\text{2}$Brainnetome Center, Institute of Automation, Chinese Academy of Sciences. \\
$^\text{3}$Beijing Key Laboratory of Brainnetome and Brain-Computer Interface, Institute of Automation, Chinese Academy of Sciences, Beijing, China.\\
correspondence author {\Letter}: \email{jiangtz@nlpr.ia.ac.cn} \\
$^\text{4}$DAMO Academy, Alibaba Group, China $^\text{5}$Hupan Lab, Hangzhou,  China \\
}

\maketitle 

\begin{abstract}

Spatial proteomics maps protein distributions in tissues, providing transformative insights for life sciences. However, current sequencing-based technologies suffer from low spatial resolution, and substantial inter-tissue variability in protein expression further compromises the performance of existing molecular data prediction methods. In this work, we introduce the novel task of spatial super-resolution for sequencing-based spatial proteomics (seq-SP) and, to the best of our knowledge, propose the first deep learning model for this task—Neural Proteomics Fields (NPF). NPF formulates seq-SP as a protein reconstruction problem in continuous space by training a dedicated network for each tissue. The model comprises a Spatial Modeling Module, which learns tissue-specific protein spatial distributions, and a Morphology Modeling Module, which extracts tissue-specific morphological features. Furthermore, to facilitate rigorous evaluation, we establish an open-source benchmark dataset, Pseudo-Visium SP, for this task. Experimental results demonstrate that NPF achieves state-of-the-art performance with fewer learnable parameters, underscoring its potential for advancing spatial proteomics research. Our code and dataset are publicly available at \url{https://github.com/Bokai-Zhao/NPF}.


\end{abstract}

\section{Introduction}

Spatial omics technologies, which profile molecular features in situ, are redefining biological discovery\cite{moffitt2022emerging}. In particular, spatial proteomics (SP), which maps protein distributions in intact tissues, was named Nature Methods’ 2024 Method of the Year\cite{karimi2024method}, accelerating advances in tumor microenvironment studies\cite{rozenblatt2020human} and biomolecular atlas construction\cite{jain2023advances}. However, despite these breakthroughs, current SP methods are still challenged by fundamental trade-offs.

\begin{figure}[t]
\includegraphics[width=\textwidth]{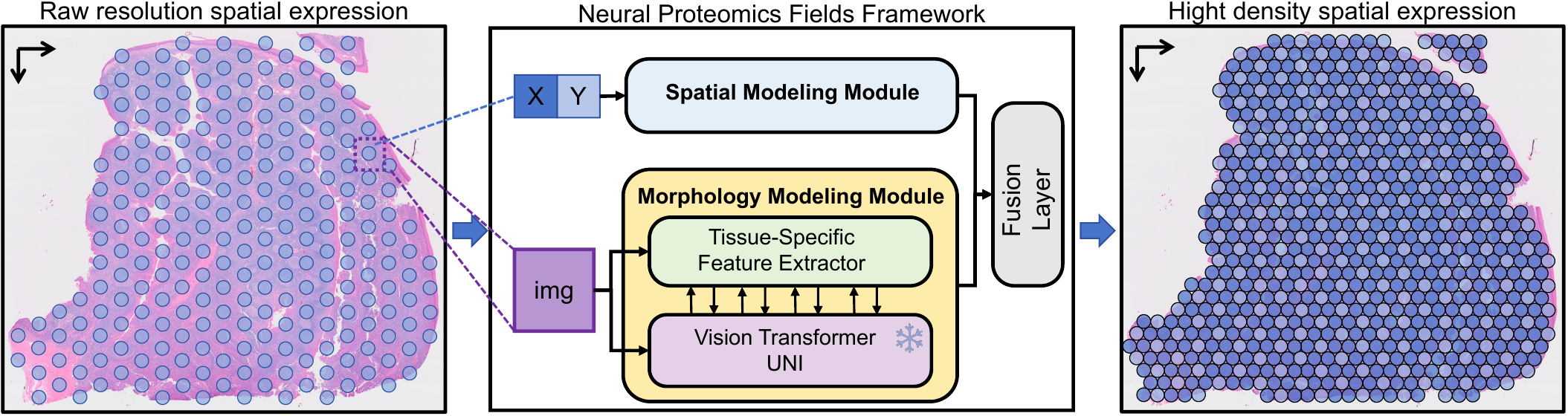}
\caption{Overview of NPF. The model takes spatial coordinates of sampling spots and corresponding image patches from whole-slide images(WSI) as input. Trained on sparsely sampled SP data, NPF predicts protein expression at unsampled locations via continuous spatial reconstruction by its implicit neural representations.} \label{fig_1}
\end{figure}

Current SP technologies can be broadly divided into two categories, each with its own limitations. Conventional imaging‐based approaches\cite{giesen2014highly,angelo2014multiplexed,goltsev2018deep}—limited by the finite number of metal isotopes or fluorescent channels—restrict the multiplexing capacity of protein detection. In recent years, to overcome this throughput bottleneck, next‐generation sequencing (NGS) methods have been adopted for SP. These sequencing‐based techniques (seq-SP) profile high-plex protein expression by sampling discrete spots on tissue sections mounted on glass slides (as shown in Fig.~\ref{fig_1}). However, the spacing and size of these spots impose a significant constraint on spatial resolution. While employing more advanced hardware to boost resolution would substantially increase sequencing costs, emerging deep learning approaches offer a promising alternative by predicting protein expression at unsampled locations—a process we refer to as spatial super-resolution for seq-SP, an area where research remains very limited\cite{hu2025high}.

The challenge of limited spatial resolution is not unique to seq-SP but also affects spatial transcriptomics (ST). Recent ST spatial super-resolution work\cite{shi2024high} has provided valuable insights: for example, STNet\cite{he2020integrating} and istar\cite{zhang2024inferring} rely solely on histological RGB patches to predict gene expression, while STAGE\cite{li2024high} leverages neighboring spot information.
Although these approaches offer useful inspiration, adapting them for protein expression prediction is nontrivial. Protein expression exhibits substantial inter-tissue variability. Even slides from the same organ may display remarkably different distributions due to factors such as age, sex, and lifestyle. Likewise, the relationship between tissue morphology and protein expression can be highly tissue-specific. Furthermore, the field lacks standardized benchmarks; the scarcity of publicly available datasets and limited sample sizes pose challenges for evaluating and comparing existing methods.

To address these challenges, we propose Neural Proteomics Fields (NPF), which, to the best of our knowledge, is the first deep learning method designed specifically for seq-SP spatial super-resolution. Inspired by Neural Radiance Fields (NeRF) \cite{mildenhall2021nerf}, NPF formulates SP prediction as a protein reconstruction problem in continuous space. By training a dedicated network for each tissue slice, NPF can effectively capture the unique spatial distribution of proteins and its relationship with histological morphology. Specifically, NPF comprises two modules: the Spatial Modeling Module(SMM) and the Morphology Modeling Module(MMM). The Spatial Modeling Module learns tissue-specific protein spatial distributions from spatial coordinates. The Morphology Modeling Module features two branches: one employs a frozen pathology foundation model (UNI\cite{chen2024towards}) to extract general morphological features, while the other, the Tissue-Specific Feature Extractor, works synergistically with UNI to capture the unique relationship between tissue morphology and proteomics for each tissue. In this way, these branches collaboratively facilitate the efficient learning of tissue-specific morphological characteristics, while simultaneously mitigating overfitting.

Furthermore, to promote a comprehensive and fair evaluation of seq-SP spatial super-resolution methods, we introduce the first open-source benchmark dataset, Pseudo-Visium SP. Constructed using a high-resolution multi-fluorescence imaging-based virtual spot generation method, this dataset simulates the spatial distribution of the 10X Genomics platform and supports rigorous cross-validation. We benchmarked NPF against representative methods in spatial transcriptomics\cite{he2020integrating,zhang2024inferring} and other predictive approaches on both the Pseudo-Visium SP dataset and real-world data from 10X Visium\cite{10xgenomics}. Experimental results demonstrate that NPF achieves state-of-the-art performance with substantially fewer learnable parameters, underscoring its potential for advancing SP research.

Our key contributions are summarized as follows:

\begin{itemize}
\item We introduce the novel task of spatial super-resolution for seq-SP and, to the best of our knowledge, propose the first deep learning method, Neural Proteomics Fields (NPF), which formulates the task as a protein reconstruction problem in continuous space, thus effectively handle inter-tissue variability via capturing each tissue’s unique protein distribution.

\item We propose a Spatial Modeling Module, which captures tissue-specific protein spatial patterns, significantly enhancing predictive performance.
\item We also develop a Morphology Modeling Module, which efficiently captures tissue-specific morphological features.

\item We establish the first open-source benchmark dataset for seq-SP spatial super-resolution, accompanied by standardized evaluation protocols.
\end{itemize}

\section{Methodology}
\begin{figure}[t]
\includegraphics[width=\textwidth]{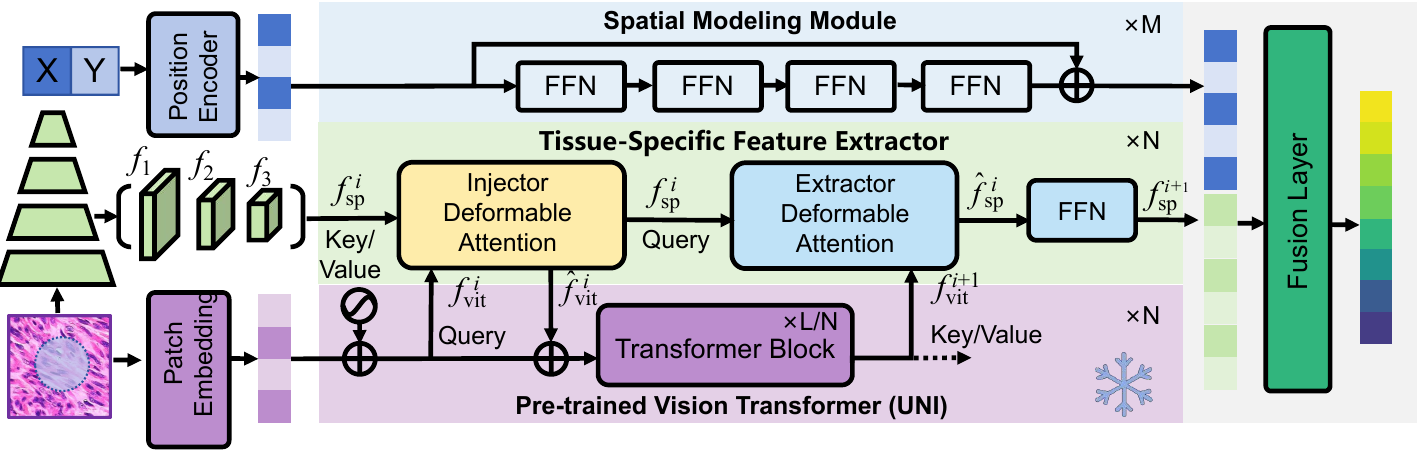}
\caption{Schematic representation of the NPF framework. NPF jointly modeling the spatial relationships and multi-scale pathological features using a dual-branch architecture. The core of the framework includes: (1) Spatial Modeling Module (SMM) inspired by NeRF\cite{mildenhall2021nerf}, which maps discrete coordinates to continuous high-dimensional space representations; (2)Morphology Modeling Module, which integrates features from a frozen pathology foundation model (UNI\cite{chen2024towards}) and a dedicated Tissue-Specific Feature Extractor (TSFE). After fusion of spatial representations and image features, the target protein expression is decoded through an MLP, enabling end-to-end SP prediction.  } \label{fig_2}
\end{figure}

\subsection{Problem Formulation}

By drawing an analogy to 3D scene reconstruction in computer vision, the problem of improving SP resolution can be viewed as generating continuous protein expression from discrete protein samples. In the context of 3D scene reconstruction, NeRF\cite{mildenhall2021nerf} optimizes an underlying continuous volumetric scene function through a sparse set of input views. This approach represents scenes using deep neural networks.

In the SP prediction task, we treat discrete spot protein samples as sparse views, using a deep neural network to represent each sample. Our NPF framework takes the spatial coordinates $(x, y)$ of the sampling spots and the fine-grained, multi-layered tissue image representation extracted from corresponding tissue images as input. This approach enables the model to generate a dense protein expression map (Fig. \ref{fig_1}), thereby enhancing the spatial resolution of seq-SP.

In NPF, we use  $I \in R^{ 3 \times H\times W }$ to represent a spot image from WSI, $H$, $W$ represents the image height and width. The observed protein expressions in each spot are denoted as $P = \{p_1,p_2,...,p_k  \}$ where $k$ is the total number of proteins. Our goal is to minimize the mean squared error (MSE) between $P' = NPF ( I,\ [x,y] \ | \ \theta) $ and $P$ by optimizing the network parameters$\theta$.

\subsection{Neural Proteomics Fields Framework}

\textbf{Overview.} The component-level Schematic of our method is illustrated in Fig.\ref{fig_2}. NPF employs a dual-branch architecture to capture spatial relationships and tissue morphological features, comprising a Spatial Modeling Module and a Morphology Modeling Module.
Detailed descriptions of each component involved in NPF are provided in the following sections.

\subsubsection{Spatial Modeling Module.}
To overcome the discretization limitation inherent in SP data, we propose a continuous spatial encoding scheme through differentiable coordinate transformation. The core innovation lies in a frequency-aware position encoder that projects normalized 2D coordinates $(x,y)$ into a continuous high-dimensional manifold using spectral embedding:

\begin{equation}\label{eq:positional_encoding}
\gamma(p) = \bigoplus_{l=0}^{L-1} \left[\sin(2^l\pi p), \cos(2^l\pi p)\right]
\end{equation}

The encoded features undergo hierarchical refinement via $M$ cascaded MLP blocks with residual connections, progressively transforming the initial $2L$-dimensional embeddings into an $n$-dimensional latent representation ($n=1024$ in our implementation). This architecture learns implicit spatial continuity, enabling systematic identification of protein expression gradients across tissue microenvironment.
In this study, we set $L=6$, $M=3$.

\subsubsection{Morphology Modeling Module.}
Taking inspiration from ViT-Adapter\cite{chen2022vision}, our MMM incorporates a frozen pre-trained pathology foundation model (UNI\cite{chen2024towards}) that learns general feature, along with a Tissue-Specific Feature Extractor that learns tissue-specific features from specific WSI. 

\noindent\textbf{\textit{Pathology-Foundation Model.}} UNI\cite{chen2024towards} is a standard ViT-Large\cite{dosovitskiy2020image} pre-trained on 100 million pathological images using DINOv2\cite{oquab2023dinov2}. The input image is initially processed by a patch embedding layer, which splits the image into $16 \times 16$  patches. A learnable \texttt{[CLS]} token is prepended to the patch sequence, followed by the addition of trainable positional embeddings. These tokens added with the position embedding, are processed through $L$ encoder layers ($L=24$).

\noindent\textbf{\textit{Tissue-Specific Feature Extractor.}} The input image is first passed through a CNN-based pyramid convolutional network\cite{lin2017feature} to extract multi-resolution features. Three target resolutions ($1/8$, $1/16$, and $1/32$) are used to gather $D$-dimensional spatial features. These feature maps are flattened and concatenated into the input $f^1_{sp}$ for feature interaction.

We divide the encoder of ViT into $N=4$ blocks, each containing $L/N=6$ encoder layers. For the $i$-th block, we inject $f^i_{sp}$ into the input feature $f^i_{vit}$ (without the \texttt{[CLS]} token) using cross-attention, as shown in Equation \ref{e2}.
\begin{equation}\label{e2}
\hat{f}^i_{vit} = f^i_{vit} + \mathrm{CrossAttn}(\mathrm{norm}(f^i_{vit}),\mathrm{norm}(f^i_{sp})),
\end{equation}
The feature of  $i$-th ViT block $\hat{f}^i_{vit}$ is then passed to the next block and get  $f^{i+1}_{vit} = \mathrm{ViT}(\hat{f}^i_{vit})$.

Next, we apply a module with cross-attention and FFN to extract fine-grained features:

\begin{equation}
\hat{f}^{i}_{sp} = f^{i}_{sp} + \mathrm{CrossAttn}(\mathrm{norm}(f^i_{sp}),\mathrm{norm}(f^{i+1}_{vit})) ,
\end{equation}

\begin{equation}
f^{i+1}_{sp} = \hat{f}^{i}_{sp} + \mathrm{FFN}(\mathrm{norm}(\hat{f}^i_{sp}) ),
\end{equation}

Here, the cross-attention layer uses Deformable Attention\cite{zhu2020deformable}. Finally, we apply mean pooling to get the final histological feature.

\subsection{Spatial Proteomics Prediction Benchmark Dataset}

Real seq-SP employs sparse sampling, which inherently limits dense prediction evaluation capabilities. To overcome this constraint, we developed the Pseudo-Visium SP dataset through virtual sampling augmentation of publicly available glioblastoma CODEX data \cite{greenwald2024integrative}, generating densely distributed spatial proteomic profiles. Our virtual sampling method preserved spatial correlation structures while enabling dense evaluation protocols.

\begin{figure}[t]
\includegraphics[width=\textwidth]{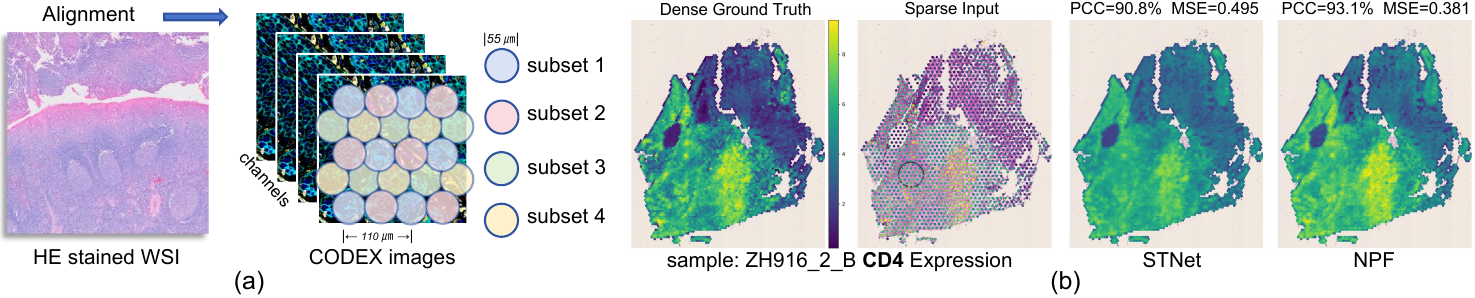}
\caption{(a) Schematic diagram of constructing Pseudo-Visium SP datasets. (b) Visualization of the prediction results of STNet and NPF on CD4 protein expression. } \label{fig_3}
\end{figure}

\noindent\textbf{Pseudo-Visium SP.} The CODEX dataset\cite{greenwald2024integrative} (sample=12) includes 40 imaging channels representing the distribution of each protein. As illustrated in Fig.~\ref{fig_3}, our preprocessing pipeline comprised:
1) {Image registration}: Rigid alignment of H\&E-stained images with CODEX maximum-intensity projections using ANTs \cite{avants2009advanced};
2) {Virtual spot generation}: Hexagonal grid placement (100$\mu$m inter-spot spacing) covering the entire tissue section;
3) {Protein quantification}: Per-channel intensity integration within 55$\mu$m-radius circular ROIs, scaled by simulated capture efficiency: $\hat{P}_i = \eta \sum I (x,y),\ x,y\in ROI_i (\eta=0.1)$
4) {Data augmentation}: Three systematic spatial shifts generating four mutually exclusive datasets for 4-fold cross-validation. 

\noindent\textbf{10X Visium data.} For real data validation, we used two publicly available SP datasets of Human Tonsil stained with HE\cite{sp-human-tonsil,sp-human-tonsil-add}, provided by 10x Genomics, which combines ST with high-plex SP using NGS on the same tissue section.

\section{Experiments}

\noindent\textbf{Implementation Details.}
For each sample, an NPF model is trained for prediction. Inputs include  $224\times224$ pixel spot images centered on coordinates normalized to the WSI dimensions (width/height).
Protein expression values are log-transformed. Training employs the Adam optimizer (initial LR=0.001) with cosine annealing scheduling and linear warmup (start LR=$10^{-6}$, epoch=5), using a batch size of 32 for 100 epochs.

\begin{table}[t]\centering
\setlength{\tabcolsep}{2pt}
\caption{Performance comparisons on Pseudo-Visium SP datasets. For each sample, we performed 4-fold cross-validation test: 70\% of spots in a subset were used for training and 30\% for validation, with testing on three held-out subsets.}\label{tab1}
\begin{tabular}{ccc|cccccccc}
\hline
\multicolumn{2}{c|}{\multirow{2}{*}{\begin{tabular}[c]{@{}c@{}}Dataset\\ sample\end{tabular}}} & \multirow{3}{*}{Metric} & \multicolumn{8}{c}{Method}                                                                                                                                                                                                      \\ \cline{4-11} 
\multicolumn{2}{c|}{}                                                                          &                         & \multicolumn{2}{c|}{Interpolate}                           & \multicolumn{3}{c|}{Image Feature}                                                    & \multicolumn{2}{c|}{ST-based}                            & \textbf{ours}   \\ \cline{1-2} \cline{4-11} 
\multicolumn{1}{c|}{Subject}                         & \multicolumn{1}{c|}{n}                 &                         & \multicolumn{1}{c|}{Nearest} & \multicolumn{1}{c|}{KNN}    & \multicolumn{1}{c|}{Res50} & \multicolumn{1}{c|}{ViT-B} & \multicolumn{1}{c|}{Swin-T} & \multicolumn{1}{c|}{istar} & \multicolumn{1}{c|}{STNet}  & \textbf{NPF}    \\ \hline
\multicolumn{3}{c|}{learnable params}                                                                                    & \multicolumn{1}{c|}{-}       & \multicolumn{1}{c|}{-}      & \multicolumn{1}{c|}{25.6M} & \multicolumn{1}{c|}{85.8M} & \multicolumn{1}{c|}{27.6M}  & \multicolumn{1}{c|}{0.4M}  & \multicolumn{1}{c|}{7.98M}  & 25.5M           \\ \hline
\multicolumn{1}{c|}{\multirow{2}{*}{MGH258}}        & \multicolumn{1}{c|}{\multirow{2}{*}{1}}  & MSE↓                    & 0.5212                       & \multicolumn{1}{c|}{0.3316} & 0.1594                     & 0.1955                     & \multicolumn{1}{c|}{\underline{0.1544}} & 0.3196                     & \multicolumn{1}{c|}{0.1651} & \textbf{0.1271}          \\

\multicolumn{1}{c|}{}                               & \multicolumn{1}{c|}{}                    & PCC↑                    & 0.6098                       & \multicolumn{1}{c|}{0.6924} & 0.8434                     & 0.8042                     & \multicolumn{1}{c|}{\underline{0.8538}} & 0.7292                     & \multicolumn{1}{c|}{0.8381} & \textbf{0.8789}          \\ 
\multicolumn{1}{c|}{\multirow{2}{*}{ZH1041}}        & \multicolumn{1}{c|}{\multirow{2}{*}{1}}  & MSE↓                    & 0.9480                       & \multicolumn{1}{c|}{0.4878} & 0.2537                     & 0.4033                     & \multicolumn{1}{c|}{\underline{0.2455}} & 0.6397                     & \multicolumn{1}{c|}{0.2704} & \textbf{0.1790}          \\
\multicolumn{1}{c|}{}                               & \multicolumn{1}{c|}{}                    & PCC↑                    & 0.5140                       & \multicolumn{1}{c|}{0.6680} & 0.8171                     & 0.7043                     & \multicolumn{1}{c|}{\underline{0.8330}} & 0.6301                     & \multicolumn{1}{c|}{0.8104} & \textbf{0.8746}          \\
\multicolumn{1}{c|}{\multirow{2}{*}{ZH1007}}        & \multicolumn{1}{c|}{\multirow{2}{*}{2}}  & MSE↓                    & 0.6455                       & \multicolumn{1}{c|}{0.3852} & \underline{0.2268}                     & 0.4141                     & \multicolumn{1}{c|}{0.2379} & 0.3811                     & \multicolumn{1}{c|}{0.2424} & \textbf{0.1529}          \\
\multicolumn{1}{c|}{}                               & \multicolumn{1}{c|}{}                    & PCC↑                    & 0.6271                       & \multicolumn{1}{c|}{0.7224} & \underline{0.8467}                     & 0.7064                     & \multicolumn{1}{c|}{0.8427} & 0.7431                     & \multicolumn{1}{c|}{0.8352} & \textbf{0.8923}          \\ 
\multicolumn{1}{c|}{\multirow{2}{*}{ZH1019}}        & \multicolumn{1}{c|}{\multirow{2}{*}{2}}  & MSE↓                    & 0.5851                       & \multicolumn{1}{c|}{0.3461} & \underline{0.2231}                     & 0.3436                     & \multicolumn{1}{c|}{0.2281} & 0.3652                     & \multicolumn{1}{c|}{0.2579} & \textbf{0.1826}          \\
\multicolumn{1}{c|}{}                               & \multicolumn{1}{c|}{}                    & PCC↑                    & 0.5383                       & \multicolumn{1}{c|}{0.6557} & \underline{0.7716}                     & 0.6259                     & \multicolumn{1}{c|}{0.7664} & 0.6656                     & \multicolumn{1}{c|}{0.7222} & \textbf{0.8191}          \\ 
\multicolumn{1}{c|}{\multirow{2}{*}{ZH811}}         & \multicolumn{1}{c|}{\multirow{2}{*}{3}}  & MSE↓                    & 0.6261                       & \multicolumn{1}{c|}{0.3780} & \underline{0.2022}                     & 0.4079                     & \multicolumn{1}{c|}{0.2228} & 0.3799                     & \multicolumn{1}{c|}{0.2311} & \textbf{0.1475}          \\
\multicolumn{1}{c|}{}                               & \multicolumn{1}{c|}{}                    & PCC↑                    & 0.6213                       & \multicolumn{1}{c|}{0.7184} & \underline{0.8377}                     & 0.6760                     & \multicolumn{1}{c|}{0.8157} & 0.7222                     & \multicolumn{1}{c|}{0.8060} & \textbf{0.8696}          \\ 
\multicolumn{1}{c|}{\multirow{2}{*}{ZH916}}         & \multicolumn{1}{c|}{\multirow{2}{*}{3}}  & MSE↓                    & 0.6707                       & \multicolumn{1}{c|}{0.4005} & \underline{0.2169}                     & 0.5095                     & \multicolumn{1}{c|}{0.2392} & 0.3972                     & \multicolumn{1}{c|}{0.2546} & \textbf{0.1628}          \\
\multicolumn{1}{c|}{}                               & \multicolumn{1}{c|}{}                    & PCC↑                    & 0.6669                       & \multicolumn{1}{c|}{0.7621} & \underline{0.8782}                     & 0.7139                     & \multicolumn{1}{c|}{0.8707} & 0.7632                     & \multicolumn{1}{c|}{0.8508} & \textbf{0.9042}          \\ \hline
\multicolumn{1}{c|}{\multirow{2}{*}{\textbf{Mean}}} & \multicolumn{1}{c|}{\multirow{2}{*}{12}} & MSE↓                    & 0.6517                       & \multicolumn{1}{c|}{0.3848} & \underline{0.2142}                     & 0.4056                     & \multicolumn{1}{c|}{0.2265} & 0.3986                     & \multicolumn{1}{c|}{0.2411} & \textbf{0.1590} \\
\multicolumn{1}{c|}{}                               & \multicolumn{1}{c|}{}                    & PCC↑                    & 0.6099                       & \multicolumn{1}{c|}{0.7132} & \underline{0.8371}                     & 0.6952                     & \multicolumn{1}{c|}{0.8303} & 0.7194                     & \multicolumn{1}{c|}{0.8112} & \textbf{0.8748} \\ \hline
\end{tabular}
\end{table}

\begin{table}[t]\centering
\setlength{\tabcolsep}{3pt}
\caption{Performance of NPF versus top baselines(Res50, STNet) on 10X Visium SP data with a 7:1:2 train/validation/test split.}\label{tab2}
\begin{tabular}{c|cc|cc|cc}
\hline
\multirow{2}{*}{\begin{tabular}[c]{@{}c@{}}10X Visium CytAssist \\ Gene and Protein Expression\end{tabular}} & \multicolumn{2}{c|}{Res50}         & \multicolumn{2}{c|}{STNet}         & \multicolumn{2}{c}{\textbf{NPF}}            \\ \cline{2-7} 
                                                                                                             & \multicolumn{1}{c|}{MSE↓} & PCC↑   & \multicolumn{1}{c|}{MSE↓} & PCC↑   & \multicolumn{1}{c|}{MSE↓} & PCC↑   \\ \hline
Human Tonsil                                                                                                 & 0.1241                    & 0.7247 & 0.1168                    & 0.7316 & \textbf{0.0648}                    & \textbf{0.8125} \\ \hline
Human Tonsil Add-on Antibodies                                                                               & 0.1588                    & 0.6007 & 0.1572                    & 0.6140 & \textbf{0.1184}                    & \textbf{0.7104} \\ \hline
\end{tabular}
\end{table}

\subsection{Performance Comparison}

\textbf{Cross-Validation on Pseudo-Visium SP.}
We benchmark NPF against three methodological categories:
1) Image feature based approaches: ResNet50 \cite{he2016deep} (Res50), Swin Transformer \cite{liu2021swin} (Swin-T), and Vision Transformer \cite{dosovitskiy2020image} (ViT-B) for image feature extraction;
3) ST predictors, i.e.istar\cite{zhang2024inferring} and STNet\cite{he2020integrating}.
To streamline result presentation, we aggregate samples into six subject-based subsets, reporting mean metrics per subset. Every subset has $n$ samples.
As shown in Table~\ref{tab1}, NPF consistently outperformed all baselines across subsets, achieving a minimum Pearson Correlation Coefficient (PCC) improvement of 3.8\%, and an MSE reduction of 0.06. Notably, NPF attain these results with nearly equivalent learnable parameters to ResNet50, demonstrating enhanced parameter efficiency.

\noindent\textbf{Evaluation on 10X Visium SP data.} We further benchmark NPF against top methods (ResNet50, STNet) on two real-world SP datasets. As shown in Table~\ref{tab2}, NPF achieves 8.1\%/9.6\% PCC improvements and 0.05/0.04 MSE reductions over best baselines, demonstrating enhanced robustness and cross-tissue generalization capability.

\noindent\textbf{Qualitative Analysis.} Visualization results demonstrate NPF's superior accuracy in protein expression prediction. Owing to space constraints, Fig.~\ref{fig_3} provides a representative comparison between STNet and NPF using the Pseudo-Visium SP dataset, highlighting enhanced spatial fidelity in microdomain resolution.

\subsection{Ablation Studies}\label{sec:ae}

\noindent\textbf{Ablation Study on Spatial Modeling Effectiveness.}
We evaluated spatial modeling through systematic integration of SMM with diverse image encoders (ResNet50, Swin-T, ViT) on the Pseudo-Visium SP benchmark. As quantified in Table~\ref{tab3}, SMM consistently enhanced prediction accuracy across architectures, achieving PCC improvements of 1.8\%, 2.5\%, and 12.9\% with concurrent MSE reductions of 0.03, 0.04, and 0.18 respectively. Notably, ViT+SMM demonstrated the most significant gains ($\Delta$PCC=+12.9\%), underscoring synergistic coupling between spatial modeling and tissue-specific morphological features extraction.
This empirical validation reveals that explicit spatial modeling enables effective utilization of global spatial protein gradients, thereby advancing SP prediction through position feature learning.

\noindent\textbf{Ablation Study on NPF Architecture. }
We conduct ablation experiments on core modules of NPF. Table~\ref{tab4} shows that all three proposed modules play indispensable roles, proving their effectiveness. Furthermore, we find that TSFE, in particular, significantly impacts performance,  followed by UNI, highlighting their key and mutually enhancing enhancement.
We attribute this to TSFE extracting tissue-specific morphological features, while UNI extracts general ones, enabling the model to effectively focus on and merge both general and specific level features, thereby enhancing the model's ability to mine tissue-rich information and accurately predict protein expression.

The performance hierarchy (Full NPF > Any Dual-Module > Single-Module) establishes the necessity of synergistic integration between three modules.

\begin{table}[t]\centering
\setlength{\tabcolsep}{6pt}
\caption{Ablation study on spatial modeling with Pseudo-Visium SP dataset.}\label{tab3}
\begin{tabular}{c|ccc|ccc|c}
\hline
Method & \multicolumn{3}{c|}{only Image Feature}                                & \multicolumn{3}{c|}{with SMM}                                   & \textbf{ours}            \\ \hline
Metric & \multicolumn{1}{c|}{Res50} & \multicolumn{1}{c|}{Swin-T} & ViT-B  & \multicolumn{1}{c|}{Res50} & \multicolumn{1}{c|}{Swin-T} & ViT-B & \textbf{NPF}             \\ \hline
MSE↓   & 0.2142                     & 0.2265                      & 0.4056 & 0.1826                         & 0.1904                          & 0.2261    & \textbf{0.1590} \\ \hline
PCC↑   & 0.8371                     & 0.8303                      & 0.6952 & 0.8547                         & 0.8555                          & 0.8238    & \textbf{0.8748} \\ \hline
\end{tabular}
\end{table}

\begin{table}[t]\centering
\setlength{\tabcolsep}{6pt}
\caption{Ablation study on NPF architecture with Pseudo-Visium SP.}\label{tab4}

\begin{tabular}{c|ccc|ccc|c}
\hline
module & \multicolumn{3}{c|}{Single-Module} & \multicolumn{3}{c|}{Dual-Module} & \textbf{ours}            \\ \hline
SMM    &  \checkmark             &              &              &  \checkmark             &              &  \checkmark            &  \checkmark               \\
TSFE   &               &  \checkmark            &              &  \checkmark             &  \checkmark            &              &  \checkmark               \\
UNI    &               &              &  \checkmark            &               &  \checkmark            &  \checkmark            &  \checkmark               \\ \hline
MSE↓   & 0.5615        & 0.3005       & 0.2683       & 0.2389        & 0.2055       & 0.2490       & \textbf{0.1590} \\ \hline
PCC↑   & 0.5169        & 0.7663       & 0.7891       & 0.8093        & 0.8404       & 0.8017       & \textbf{0.8748} \\ \hline
\end{tabular}
\end{table}
\section{Summary}

We introduce the novel task of spatial super-resolution for seq-SP and propose NPF, an implicit neural framework designed for SP prediction from WSI. This framework integrates spatial modeling with a Morphology Modeling Module, providing a powerful approach for high-plex protein expression prediction. Extensive experiments on pseudo and real-world datasets demonstrate NPF’s state-of-the-art performance. Notably, NPF establishes a new paradigm for spatial modeling in spatial omics. Future work includes broadening NPF’s applications, utilizing larger datasets, and integrating with other spatial omics for deeper insights into tissue complexity.

\begin{credits}
\subsubsection{\ackname} This work was partially supported by STI2030-Major Projects (No. 2021ZD0200200) and the Equipment Development Project of the Chinese Academy of Sciences (YJKYYQ20190040).

\subsubsection{\discintname}
The authors declare that they have no competing interests.
\end{credits}

\bibliographystyle{splncs04}
\bibliography{Paper-1305}

\begin{thebibliography}{10}
\providecommand{\url}[1]{\texttt{#1}}
\providecommand{\urlprefix}{URL }
\providecommand{\doi}[1]{https://doi.org/#1}

\bibitem{sp-human-tonsil}
Human tonsil, \url{https://www.10xgenomics.com/datasets/gene-protein-expression-library-of-human-tonsil-cytassist-ffpe-2-standard}

\bibitem{sp-human-tonsil-add}
Human tonsil add-on antibodies, \url{https://www.10xgenomics.com/datasets/visium-cytassist-gene-and-protein-expression-library-of-human-tonsil-with-add-on-antibodies-h-e-6-5-mm-ffpe-2-standard}

\bibitem{10xgenomics}
{10x Genomics}: Spatial gene and protein expression, \url{www.10xgenomics.com}

\bibitem{angelo2014multiplexed}
Angelo, M., Bendall, S.C., Finck, R., Hale, M.B., Hitzman, C., Borowsky, A.D., Levenson, R.M., Lowe, J.B., Liu, S.D., Zhao, S., et~al.: Multiplexed ion beam imaging of human breast tumors. Nature medicine  \textbf{20}(4),  436--442 (2014)

\bibitem{avants2009advanced}
Avants, B.B., Tustison, N., Song, G., et~al.: Advanced normalization tools (ants). Insight j  \textbf{2}(365),  1--35 (2009)

\bibitem{chen2024towards}
Chen, R.J., Ding, T., Lu, M.Y., Williamson, D.F., Jaume, G., Song, A.H., Chen, B., Zhang, A., Shao, D., Shaban, M., et~al.: Towards a general-purpose foundation model for computational pathology. Nature Medicine  \textbf{30}(3),  850--862 (2024)

\bibitem{chen2022vision}
Chen, Z., Duan, Y., Wang, W., He, J., Lu, T., Dai, J., Qiao, Y.: Vision transformer adapter for dense predictions. arXiv preprint arXiv:2205.08534  (2022)

\bibitem{dosovitskiy2020image}
Dosovitskiy, A., Beyer, L., Kolesnikov, A., Weissenborn, D., Zhai, X., Unterthiner, T., Dehghani, M., Minderer, M., Heigold, G., Gelly, S., et~al.: An image is worth 16x16 words: Transformers for image recognition at scale. arXiv preprint arXiv:2010.11929  (2020)

\bibitem{giesen2014highly}
Giesen, C., Wang, H.A., Schapiro, D., Zivanovic, N., Jacobs, A., Hattendorf, B., Sch{\"u}ffler, P.J., Grolimund, D., Buhmann, J.M., Brandt, S., et~al.: Highly multiplexed imaging of tumor tissues with subcellular resolution by mass cytometry. Nature methods  \textbf{11}(4),  417--422 (2014)

\bibitem{goltsev2018deep}
Goltsev, Y., Samusik, N., Kennedy-Darling, J., Bhate, S., Hale, M., Vazquez, G., Black, S., Nolan, G.P.: Deep profiling of mouse splenic architecture with codex multiplexed imaging. Cell  \textbf{174}(4),  968--981 (2018)

\bibitem{greenwald2024integrative}
Greenwald, A.C., Darnell, N.G., Hoefflin, R., Simkin, D., Mount, C.W., Castro, L.N.G., Harnik, Y., Dumont, S., Hirsch, D., Nomura, M., et~al.: Integrative spatial analysis reveals a multi-layered organization of glioblastoma. Cell  \textbf{187}(10),  2485--2501 (2024)

\bibitem{he2020integrating}
He, B., Bergenstr{\aa}hle, L., Stenbeck, L., Abid, A., Andersson, A., Borg, {\AA}., Maaskola, J., Lundeberg, J., Zou, J.: Integrating spatial gene expression and breast tumour morphology via deep learning. Nature biomedical engineering  \textbf{4}(8),  827--834 (2020)

\bibitem{he2016deep}
He, K., Zhang, X., Ren, S., Sun, J.: Deep residual learning for image recognition. In: Proceedings of the IEEE conference on computer vision and pattern recognition. pp. 770--778 (2016)

\bibitem{hu2025high}
Hu, B., He, R., Pang, K., Wang, G., Wang, N., Zhu, W., Sui, X., Teng, H., Liu, T., Zhu, J., et~al.: High-resolution spatially resolved proteomics of complex tissues based on microfluidics and transfer learning. Cell  (2025)

\bibitem{jain2023advances}
Jain, S., Pei, L., Spraggins, J.M., Angelo, M., Carson, J.P., Gehlenborg, N., Ginty, F., Gon{\c{c}}alves, J.P., Hagood, J.S., Hickey, J.W., et~al.: Advances and prospects for the human biomolecular atlas program (hubmap). Nature cell biology  \textbf{25}(8),  1089--1100 (2023)

\bibitem{karimi2024method}
Karimi, E., Simo, N., Milet, N., TE, W., ALSH, A., QU, N., AIL, L., ABS, R., ALIND, A., GOODMA, N.M., et~al.: Method of the year 2024: spatial proteomics. Nat Methods  \textbf{21},  2195--2196 (2024)

\bibitem{li2024high}
Li, S., Gai, K., Dong, K., Zhang, Y., Zhang, S.: High-density generation of spatial transcriptomics with stage. Nucleic Acids Research  \textbf{52}(9),  4843--4856 (2024)

\bibitem{lin2017feature}
Lin, T.Y., Doll{\'a}r, P., Girshick, R., He, K., Hariharan, B., Belongie, S.: Feature pyramid networks for object detection. In: Proceedings of the IEEE conference on computer vision and pattern recognition. pp. 2117--2125 (2017)

\bibitem{liu2021swin}
Liu, Z., Lin, Y., Cao, Y., Hu, H., Wei, Y., Zhang, Z., Lin, S., Guo, B.: Swin transformer: Hierarchical vision transformer using shifted windows. In: Proceedings of the IEEE/CVF international conference on computer vision. pp. 10012--10022 (2021)

\bibitem{mildenhall2021nerf}
Mildenhall, B., Srinivasan, P.P., Tancik, M., Barron, J.T., Ramamoorthi, R., Ng, R.: Nerf: Representing scenes as neural radiance fields for view synthesis. Communications of the ACM  \textbf{65}(1),  99--106 (2021)

\bibitem{moffitt2022emerging}
Moffitt, J.R., Lundberg, E., Heyn, H.: The emerging landscape of spatial profiling technologies. Nature Reviews Genetics  \textbf{23}(12),  741--759 (2022)

\bibitem{oquab2023dinov2}
Oquab, M., Darcet, T., Moutakanni, T., Vo, H., Szafraniec, M., Khalidov, V., Fernandez, P., Haziza, D., Massa, F., El-Nouby, A., et~al.: Dinov2: Learning robust visual features without supervision. arXiv preprint arXiv:2304.07193  (2023)

\bibitem{rozenblatt2020human}
Rozenblatt-Rosen, O., Regev, A., Oberdoerffer, P., Nawy, T., Hupalowska, A., Rood, J.E., Ashenberg, O., Cerami, E., Coffey, R.J., Demir, E., et~al.: The human tumor atlas network: charting tumor transitions across space and time at single-cell resolution. Cell  \textbf{181}(2),  236--249 (2020)

\bibitem{shi2024high}
Shi, Z., Xue, S., Zhu, F., Min, W.: High-resolution spatial transcriptomics from histology images using histosge. In: 2024 IEEE International Conference on Bioinformatics and Biomedicine (BIBM). pp. 2402--2407. IEEE (2024)

\bibitem{zhang2024inferring}
Zhang, D., Schroeder, A., Yan, H., Yang, H., Hu, J., Lee, M.Y., Cho, K.S., Susztak, K., Xu, G.X., Feldman, M.D., et~al.: Inferring super-resolution tissue architecture by integrating spatial transcriptomics with histology. Nature biotechnology  \textbf{42}(9),  1372--1377 (2024)

\bibitem{zhu2020deformable}
Zhu, X., Su, W., Lu, L., Li, B., Wang, X., Dai, J.: Deformable detr: Deformable transformers for end-to-end object detection. arXiv preprint arXiv:2010.04159  (2020)

\end{thebibliography}

\end{document}